\documentclass[letterpaper]{jpconf}
\usepackage{graphicx}
\begin{document}
\title{The Hawking signal in density-density \\ correlations in BECs}

\author{Alessandro Fabbri}

\address{Departamento de F\'isica Te\'orica and IFIC, Universidad de
Valencia-CSIC,  C. Dr. Moliner 50, 46100 Burjassot, Spain}

\ead{afabbri@ific.uv.es}

\begin{abstract}
We outline the derivation of the Hawking quanta-partner signal in correlations, and highlight
the specific application to detect it in density-density correlations in BECs.
\end{abstract}

% \section{Introduction}
%The study of fundamental interactions in nature makes it clear the existing fracture between %the microscopic quantum mechanical description of our world and the classical general %relativistic macroscopic one. One of the main challenges theoretical physicists face is to %find a self-consistent quantum framework that takes into account the gravitational interaction %as well.

%In 1966 Parker introduced a formalism to calculate the leading quantum corrections to %Einstein's theory, QFT in curved space, where quantum matter fields are coupled to a classical %gravitational field (i.e., they propagate on a curved spacetime geometry). He applied it to %cosmology, whereas eight years later Hawking used it to get his famous result on black holes %evaporation.

%The ancestors of black holes are Mitchell's and Laplace's newtonian dark stars, compact %objects with radius $r\le r_G\equiv \frac{2GM}{c^2}$ and escape velocity $\ge c$. Although %General Relativity taught us that dark stars do not exist, it confirmed the crucial role %played by the Schwarzschild radius $r_G$, the event horizon, causal frontier between the black %hole region and the exterior.

Black holes possess surprising quantum features. In the gravitational collapse of a star forming a black hole, Hawking \cite{hawking} showed that the initial quantum vacuum state of matter fields gets distorted in such a way that our measurements, restricted to the exterior region, would detect a thermal state at the temperature $T_H=\frac{\kappa}{2\pi}$,
%\begin{equation}
%T_H=\frac{\hbar\kappa}{2\pi k_B c}\
%\end{equation}
where $\kappa$ ($=1/4M$ for Schwarzschild) is the horizon's surface gravity.
A too tiny effect, unfortunately,  since for solar mass black holes $T_H\sim 10^{-7}\ K$
is 7 orders of magnitude below the cosmic microwave background temperature.

The basic mechanism explaining the Hawking effect is pair-production in the near-horizon region, with one member escaping to infinity (the positive energy {\it Hawking quanta})
and the other falling into the black hole (the negative energy {\it partner}) \cite{HawSA}. We can see this effect by looking at correlations.
The initial vacuum state contains local correlations that are transferred, in the course of the time evolution, to nonlocal correlations between the black hole and the external region (these are precisely the correlations Hawking quanta-partner).
Such correlations, lost in the tracing operation implicit in any measurement performed in the external region, have a typical form that characterizes the Hawking effect. In the gravitational case, however, their study is of purely academic interest due to the impossibility of measuring them.

A completely different framework to test the existence and the detectability of the Hawking effect was proposed by Unruh \cite{unruh}, who used
the mathematical equivalence between light propagation in a curved spacetime and sound propagation in inhomogeneous eulerian fluids to predict the production of a thermal flux of phonons whenever an acoustic horizon (i.e. a region separating subsonic and supersonic flow) is formed. The reasoning is the same as Hawking's, and rests on the fact that the nature of the effect is kinematical, i.e. it does not depend on the underlying dynamics but just on the details of wave propagation in a black-hole like geometry.

Unruh's motivation was to test the robustness of the Hawking effect against modifications of the relativistic (hydrodynamic in this case) theory at short distances, an analysis much more difficult to perform in the gravitational case (the so called transplanckian problem \cite{jacobson}).

Among the many systems proposed (see for instance \cite{blv}), Bose-Einstein condensates (BECs) offer particularly attractive experimental conditions \cite{gacz}. Indeed, the huge difference between the Hawking temperature and that of the surrounding can be here reduced to just one order of magnitude. Still, it appears difficult to separate the thermal flux of the created Hawking-Unruh phonons from that of the (bigger) background.

The situation appears more promising if we look at the signature of this effect in the correlations, exploiting the crucial (and obvious) fact that in this context the acoustic origin of the horizon does not forbid correlation measurements between the exterior and the black hole.

We shall outline here a derivation of the Hawking effect for the case of a 2D conformal scalar field $\phi$ propagating in the two-dimensional $(t,x)$ section of the BEC acoustic geometry
\begin{equation}\label{metric}
ds^2=\frac{n}{mc(x,t)}\left[ -(c^2(x,t)-v^2)dt^2 -2vdtdx +dx^2 +dy^2 +dz^2 \right]\ ,
\end{equation}
in which we consider a one-dimensional condensate with constant number density $n$ and velocity $v$ ($<0$) along $x$ ($m$ is the mass of the atoms), and the only non-trivial quantity is the speed of sound $c(x,t)$, that can be tuned by varying the atom-atom scattering length.

The formation of the acoustic black hole is described by considering an initial spatially homogeneous configuration with $c_{in}=const.>|v|$ for $t<0$ (`in' region) turning, as $t>0$, into a nonhomogeneous one with $c(x)>|v|$ ($c(x)<|v|$) for $x>0$ ($x<0$) and an acoustic horizon (defined by $c(x)+v=0$) at $x=0$ (`out' region).

Our quantum field $\hat \phi$ is prepared in the state $|in\rangle$, which is the ground state of the decomposition
\begin{equation}\label{inde}
\hat\phi= \int_0^{+\infty}dw\Big[ \hat a_w^{u,in} u_w^{u,in}+ \hat a_w^{v,in} u_w^{v,in} + h.c. \Big] \ ,
\end{equation}
where $u_w^{u(v),in}
=\frac{e^{iwu(v)^{in}}}{\sqrt{4\pi w}}$ ($u^{in}=t-\frac{x}{v+c_{in}},\ v^{in}=t+\frac{x}{c_{in}-v}$)
are the positive frequency `in' right-moving (left-moving) modes basis in the `in' region.
The modes are ortho-normalized according to the Klein-Gordon scalar product
\begin{equation}
(u_w,u_{w'})=-i\int dx\left[ u_w \left(\frac{\partial_t+v\partial_x}{c}\right)u_{w'}^{*} - u_{w'}^{*}\left(\frac{\partial_t + v\partial_x}{c}\right)u_w\right]
\end{equation}
and $|in\rangle$ satisfies the condition $\hat a_w^{u(v),in}|in\rangle=0$ for all $w$.

To find the physical consequences of the choice of $|in\rangle$ state in the `out' region we need to evolve the `in' modes and connect them with the `out' modes basis.
Moreover, being right-moving and left-moving modes decoupled we will only consider the $u$ sector, which is the revelant one for our purposes.

A positive frequency `in' mode
\begin{equation}\label{tneg}
\phi_w (t<0)
=u_w^{u,in}
=\frac{e^{-iwu^{in}}}{\sqrt{4\pi w}},\ \ u^{in}=t-\frac{x}{v+c_{in}},
\end{equation}
%$\phi (t<0) =u_w^{u,in}=\frac{e^{-iwu^{in}}}{\sqrt{4\pi w}}$, $u^{in}=t-\frac{x}{v+c_{in}}$,
is matched with a linear combination of both positive and negative frequency `out' modes,
$u_{w}^{u,out}=\frac{e^{-iw u^{out}}}{\sqrt{4\pi w}}$, $u^{out}=t-\int\frac{dx}{v+c(x)}$ (such modes are positive frequency outside the horizon ($x>0$) and negative-frequency inside ($x<0$))
%Clearly to match with $\phi (t<0)$ we need a linear combination of both positive and negative %frequency modes, that as usual we write in the form
\begin{equation}\label{phiout}
\phi_w (t>0)=\int_0^{+\infty} d\tilde w \Big[ \alpha_{w\tilde w}u_{\tilde w}^{u,out} + \beta_{w\tilde w}u_{\tilde w}^{u,out *}\Big]
\end{equation}
with Bogoliubov coefficients $\alpha$ and $\beta$. Such coefficients are given by the scalar products
\begin{eqnarray}\label{bogcoeff}
&&\alpha_{w\tilde w}=(\phi_w,u_{\tilde w}^{u, out})=\frac{1}{2\pi}\sqrt{\frac{\tilde w}{w}}
\int \frac{dx'}{v+c(x')} e^{-i\tilde w\int^{x'}\frac{dy}{v+c(y)}+\frac{iw x'}{v+c_{in}}}, \nonumber \\ &&\beta_{w\tilde w}=-(\phi_w, u_{\tilde w}^{u, out *})=\frac{1}{2\pi}\sqrt{\frac{\tilde w}{w}}
\int \frac{dx'}{v+c(x')} e^{i\tilde w\int^{x'} \frac{dy}{v+c(y)}+\frac{iw x'}{v+c_{in}}}\ .
\end{eqnarray}
It can be shown that the global solution (\ref{tneg})-(\ref{bogcoeff})
satisfies the two-dimensional Klein-Gordon equation $\nabla^2\phi=0$ including the matching conditions at $t=0$
($[\ ]$ means variation across the jump at $t=0$)
\begin{equation}\label{boundcond}
[\phi]=0\ , \ \ \left[ \frac{(\partial_t +v\partial_x)\phi}{c} \right]=0\ ,
\end{equation}
%$[\phi]=0$, $\left[ \frac{(\partial_t +v\partial_x)\phi}{c} \right]=0$,
i.e.
\begin{eqnarray}
\frac{e^{\frac{iwx}{v+c_{in}}}}{\sqrt{w}}=\int_0^{+\infty}\frac{d\tilde w}{\sqrt{\tilde w}}
\left[ \alpha_{w\tilde w}e^{i\tilde w\int^x \frac{dy}{v+c(y)}}+
\beta_{w\tilde w}e^{-i\tilde w\int^x \frac{dy}{v+c(y)}}\right], \\
\frac{\sqrt{w}e^{\frac{iwx}{v+c_{in}}}}{v+c_{in}}=\frac{1}{v+c(x)}\int_0^{+\infty}d\tilde w \sqrt{\tilde w}
\left[ \alpha_{w\tilde w}e^{i\tilde w\int^x \frac{dy}{v+c(y)}}-
\beta_{w\tilde w}e^{-i\tilde w\int^x \frac{dy}{v+c(y)}}\right]
\end{eqnarray}
(the analysis of the four dimensional Klein-Gordon equation in the full acoustic geometry (\ref{metric}) is more involved because right-moving and left-moving modes are no more decoupled).

Eq. (\ref{phiout}) allows us to relate the operators of the `in' decomposition (\ref{inde}) with the `out' one
%(again, we consider only the $u$ sector)
\begin{equation}\label{outde}
\hat\phi= \int_0^{+\infty}d\tilde w\Big[ \hat a_{\tilde w}^{u,ext} u_{\tilde w}^{u,ext}+ \hat a_{\tilde w}^{u,bh \dagger} u_{\tilde w}^{u,bh} + \hat a_{\tilde w}^{v,out} u_{\tilde w}^{v,out} + h.c. \Big] \
\end{equation}
and in particular we get
\begin{equation}
 \hat a_{\tilde w}^{u,ext}=\int dw \Big[ \alpha_{w\tilde w}^{(x>0)}\hat a_w^{u,in}+ \beta_{w\tilde w}^{(x>0)*}\hat a_w^{u,in\dagger}\Big],
 \hat a_{\tilde w}^{u,bh\dagger}=\int dw \Big[ \alpha_{w\tilde w}^{(x<0)}\hat a_w^{u,in}+ \beta_{w\tilde w}^{(x<0)*}\hat a_w^{u,in\dagger}\Big]\ .\nonumber
 \end{equation}
Restricting to the exterior of the horizon it is not difficult to show, using the near-horizon expansion
$c(x)\sim -v+\kappa x$ with $\kappa$ the horizon's surface gravity to evaluate explicitly the main contribution to
the Bogoliubov coefficients
\begin{equation}\label{bogfuori}
\alpha_{w\tilde w}^{(x>0)}\sim \frac{1}{2\pi \kappa}\sqrt{\frac{\tilde w}{w}}\left[\frac{-iw}{\kappa(v+c_{in})}\right]^{\frac{i\tilde w}{\kappa}}\Gamma(-\frac{i\tilde w}{\kappa})\ , \ \
\beta_{w\tilde w}^{(x>0)}\sim \frac{1}{2\pi \kappa}\sqrt{\frac{\tilde w}{w}}\left[\frac{-iw}{\kappa(v+c_{in})}\right]^{-\frac{i\tilde w}{\kappa}}\Gamma(\frac{i\tilde w}{\kappa})\ ,
\end{equation}
 that at late retarded times $u$ the state $|in\rangle$ is thermally populated of Hawking-Unruh quanta
\begin{equation}
\langle in|\hat a_{\tilde w}^{u,ext\dagger}a_{\tilde w}^{u,ext}|in\rangle =\lim_{\tilde w'\to \tilde w} \int dw \beta_{w\tilde w}^{(x>0)}\beta_{w\tilde w'}^{(x>0)*}=\frac{\delta(0)}{e^{\frac{2\pi\tilde w}{\kappa }-1}}\ . \end{equation}
The Dirac $\delta$, associated to the use of continuum plane wave modes, would disappear if we considered wave-packets instead.

The possibility of exploring the interior region as well allows us to study the correlations between the emitted quanta and their partners inside the horizon. To determine the  symmetric two-point function $\langle in|\{ \hat\phi(t,x),\hat \phi(t',x')\}|in\rangle \equiv \frac{1}{2}\langle in|\hat\phi(t,x)\hat\phi(t',x')+\hat\phi(t',x')\hat\phi(t,x)|in\rangle $
%The basic quantity for our purposes is the  two-point function with one point inside
for, say, $x<0$ and $x'>0$ we need also the Bogoliubov coefficients in the interior region
\begin{eqnarray}
\label{bogdentro}
&& \alpha_{w\tilde w}^{(x<0)}\sim -\frac{1}{2\pi \kappa}\sqrt{\frac{\tilde w}{w}}\left[\frac{iw}{\kappa(v+c_{in})}\right]^{\frac{i\tilde w}{\kappa}}\Gamma(-\frac{i\tilde w}{\kappa})=
-e^{-\frac{\pi w}{\kappa}}\alpha_{w\tilde w}^{(x>0)}\ , \nonumber \\
&& \beta_{w\tilde w}^{(x<0)}\sim -\frac{1}{2\pi \kappa}\sqrt{\frac{\tilde w}{w}}\left[\frac{iw}{\kappa(v+c_{in})}\right]^{-\frac{i\tilde w}{\kappa}}\Gamma(\frac{i\tilde w}{\kappa})=
-e^{\frac{\pi w}{\kappa}}\beta_{w\tilde w}^{(x>0)} \ ,
\end{eqnarray}
and we get (we select only the relevant $u$-sector contribution)
%(the integrand is now proportional to $\alpha_{w\tilde w}^{(x<0)}\alpha_{w\tilde w'}^{(x>0)*}
%+ \beta_{w\tilde w}^{(x<0)}\beta_{w\tilde w'}^{(x>0)*}$)
%with the result
\begin{eqnarray}
&&\langle in| \{ \hat\phi (t,x),\hat\phi (t',x')\} |in\rangle =
\nonumber \\
&& \frac{1}{2}\int d\tilde  wd\tilde w' dw
\Big[ u_{\tilde w}^{u,bh}\alpha_{w\tilde w}^{(x<0)}+u_{\tilde w}^{u,bh *}\beta_{w\tilde w}^{(x<0)}\Big](t,x)
\Big[ u_{\tilde w'}^{u,ext *}\alpha_{w\tilde w'}^{(x>0)*}+u_{\tilde w'}^{u,ext }\beta_{w\tilde w'}^{(x>0)*}\Big](t',x') + c.c. \  \ \ \ \ \ \ \ \ \nonumber \\
&& = \frac{1}{2} \int d\tilde  w d\tilde w' dw u_{\tilde w}^{u,bh}(t,x)u_{\tilde w'}^{u,ext *}(t',x')
\Big[ \alpha_{w\tilde w}^{(x<0)}\alpha_{w\tilde w'}^{(x>0)*} + \beta_{w\tilde w}^{(x<0)*}\beta_{w\tilde w'}^{(x>0)}\Big] + c.c. \nonumber \\
&& = -\int \frac{d\tilde w}{4\pi\tilde w} \frac{\cos \tilde w(u^{bh}-u^{ext})}{\sinh\frac{\pi\tilde w}{\kappa}}
=-\frac{1}{4\pi} \ln\cosh \frac{\kappa(u^{bh}-u^{ext})}{2}\ ,
\label{tu}
\end{eqnarray}
where $u^{bh}=t-\int^x\frac{dy}{v+c(y)}$\ (with $x<0$) and $u^{ext}= t'-\int^{x'} \frac{dy'}{v+c(y')}$ ($x'>0$).

Going specifically to the BEC case, the quantity of interest in the experiments is the one time, normalized correlation function of the density fluctuations $G^{(2)}(t;x,x')=\frac{1}{n^2}\langle \hat n_1(t,x)\hat n_1(t',x')\rangle|_{t=t'}$. In the
hydrodynamical approximation $\hat n_1=-\frac{n}{mc^2}(\partial_t +v\partial_x)\hat \theta^1$ and the phase operator $\hat \theta_1$ is well approximated, up to a proportionality factor, by our field $\hat \phi$ ($\hat \theta^1\sim \sqrt{\frac{mc}{n}}\hat \phi$, see \cite{bafrc} for more details). If we consider points sufficiently far away so that the speed of sound is well approximated with its asymptotic values $c(x)\to c_l$, $c(x')\to c_r$  a straightforward calculation, using the crucial result (\ref{tu}), gives
\begin{equation}
G^{(2)}(t;x,x')= \frac{\kappa^2}{16\pi mn(c_rc_l)^{3/2}}\frac{c_rc_l}{(v+c_l)(v+c_r)}\frac{1}{\cosh^2[\frac{\kappa}{2}(\frac{x}{v+c_l}-\frac{x'}{v+c_r})]}\ . \end{equation}
These correlations characterize neatly the Hawking effect, namely the dependence on the surface gravity in the height and width and in particular the stationary peak at $\frac{x}{v+c_l}=\frac{x'}{v+c_r}$ identifying pairs of quanta created just inside and just outside the horizon at all times after its formation and traveling in opposite directions.

The robustness of these correlations has been numerically observed in an analysis carried out using the microscopic theory in \cite{cfrbf}, where it has been also shown that this effect is still present and clearly visible even in the presence of a (bigger) thermal background. Inserting numbers for existing experiments, we have correlations of the order of $10^{-3}$, small but not negligible. Proposals to amplify the signal by a factor 10 by following the formation of the acoustic black hole with a period of free expansion may be important for the experimental search \cite{cornell09}. These facts, together with the vast technology available nowadays, indicate that density correlation measurements are very promising for an experimental verification of the Hawking-Unruh effect in the near future.

{\bf Acknowledgements:} It is a pleasure to thank R. Balbinot, I. Carusotto and A. Recati for many useful and interesting discussions.

% (in the cosmological context, they have also been proposed to study short-distance %corrections to the correlator of temperature fluctuations in the
%CMB).

%In \cite{Balbinot:2007de}

%\begin{figure}[h] \centering \includegraphics[angle=0, height=2in] {black_hole.eps}
%\caption{...}
%\end{figure}

\end{document}